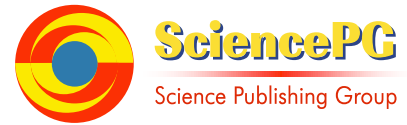

# Bile duct segmentation methods under 3D slicer applied to ERCP: advantages and disadvantages.

Abdelhadi Essamlali[1, *], Vincent Millot-Maysounabe[1], Marion Chartier[2], Grégoire Salin[2], Aymeric Becq[3], Lionel Arrivé[4], Marine Duboc Camus[2], Jérôme Szewczyk[5], Isabelle Claude[1]

[1]Biomechanics and bioengineering laboratory, University of Technology of Compiègne, Compiègne, France
[2]Digestive endoscopy department, Saint-Antoine Hospital, Paris, France
[3]Gastroenterology department, Henri Mondor Hospital, Creteil, France
[4]Radiology department, Saint-Antoine Hospital, Paris, France
[5]Institute of intelligent systems and robotics, Sorbonne University, Paris, France

**Email address:**
abdelhadi.essamlali@utc.fr (Abdelhadi Essamlali)
[*]Corresponding author

**Abstract:** This article presents an evaluation of biliary tract segmentation methods used for 3D reconstruction may be very usefull in various kind of critical interventions such as endoscopic retrograde cholangiopancreatography (ERCP) using the 3D Slicer software. This article provides an assessment of biliary tract segmentation techniques employed for 3D reconstruction, which can prove highly valuable in diverse critical procedures like endoscopic retrograde cholangiopancreatography (ERCP) through the utilization of 3D Slicer software. Three different methods, namely thresholding, flood filling, and region growing, were assessed in terms of advantages and disadvantages. The study involved 10 patient cases and employed quantitative indices and qualitative evaluation to assess the segmentations obtained by the different segmentation methods against ground truth. The results indicate that the thresholding method is almost manual and time-consuming, while the flood filling method is semi-automatic and also time-consuming. Although both methods improve segmentation quality, they are not reproducible. Therefore, an automatic method based on region growing was developed to reduce segmentation time, albeit at the expense of quality. These findings highlight the pros and cons of different conventional segmentation methods and underscore the need to explore alternative approaches such as deep learning to optimize biliary tract segmentation in the context of ERCP.

**Keywords :** Segmentation, biliary tract, MRI images, ERCP, U-Net.

## 1. Introduction

Endoscopic retrograde cholangiopancreatography (ERCP) is a procedure intended for the treatment of biliary stones (which affects 2-4% of the general population) and biliary or pancreatic neoplasia. In France, approximately 78,000 ERCP procedures are performed each year for this purpose [1]. This invasive procedure is technically complex, as it involves two distinct steps which can pose a technical challenge. The first step consists in accessing the major papilla, located in the duodenum, using a side-viewing endoscope called a duodenoscope. The second step, called biliary cannulation or catheterization, involves penetrating the bile ducts via the major papilla, using a guidewire-loaded catheter. ECRP fails in 2% of cases and adverse events are reported in 4 to 9% of cases [2]. Most challenge cases are represented by biliary malignancies, with infiltrating lesions of the biliary tree, with the risk of cholangitis and delay chemotherapy or surgery in case of wrong biliary drainage during ERCP.

Understanding the biliary anatomy is critical, but poses a significant challenge due to the complexity of the biliary tree and the interindividual anatomic variations. The success of ERCP can be hindered by a limited comprehension of biliary anatomy. This claim is supported by recent findings in an article by Becq et al [3] on this topic. While preoperative imaging techniques such as abdominal CT and cholangio-MRI images, as well as intraoperative guidance with 2D fluoroscopy provide useful information (see "Figure 1"), they do not provide a comprehensive overview of the patient biliary anatomy. Hence, segmentation and 3D reconstruction of the biliary tract based on cholangio-MRI images is an interesting prospect to overcome this limitation [3]. Reliable and robust automated segmentation could improve the accuracy and efficiency of diagnosis [4], help better plan the procedure and allow a significant time and X-ray exposure saving for patients and physicians.

However, cholangio-MRI automated segmentation faces many challenges: small target volume, uneven image brightness, low spatial resolution (especially in the inter-slice direction [5]), variable image quality, poor contrast in the intrahepatic ducts (IHDs) [6], presence of bifurcations, tortuous bile ducts, and bright gastrointestinal structures. The most important challenge remains the absence of a signal in the case of stenosis because the cholangio-MRI is a sequence that detects bile and not the walls of the bile ducts. In those conditions, manual segmentation, although feasible and robust [3], is incredibly time-consuming.

Very few studies on bile duct segmentation have been published. Ivashchenko et al. [7] employed multiscale vesselness filters from dynamic MRI images with liver-specific contrast agents to perform bile duct segmentations. The study tested the algorithm using 15 patient data sets with different types of liver malignancies. The results were evaluated by calculating the average Hausdorff distance between the central lines of the bile ducts. However, calculating the center lines of the generated segmentations is insufficient to understand the patient's bile duct anatomy, including strictures and dilatations. Some other studies employed segmentation tasks to classify specific types of diseases. For instance, Al-Oudat et al. (2021) [8] utilized the active contour method to segment the biliary tree and extract features. After feature extraction, they employed different classifiers to determine the health status of patients, whether they were normal or dilated. Although this study establishes a scientific correlation between the extracted features and the state of the biliary tree, the authors did not specify how they validated the results of their automatic segmentations.

Ralli et al. [6] conducted a study on the improvement of the quality of cholangio-MRI images and found that the regularized phase symmetry filter can enhance the bile ducts, suppress the background, and improve the image without distorting or suppressing the intrahepatic bile ducts, suggesting that it may be a better option for preprocessing biliary MRI images and achieving more accurate bile duct segmentations.

The aim of our study was to evaluate the accuracy of three classical segmentation methods compared to manual segmentation (ground truth) in terms of their respective advantages and disadvantages. These methods were applied within the specific context of cholangio-MRI image segmentation. The evaluation was conducted using quantitative and qualitative parameters, aiming to highlight the differences between these practices.

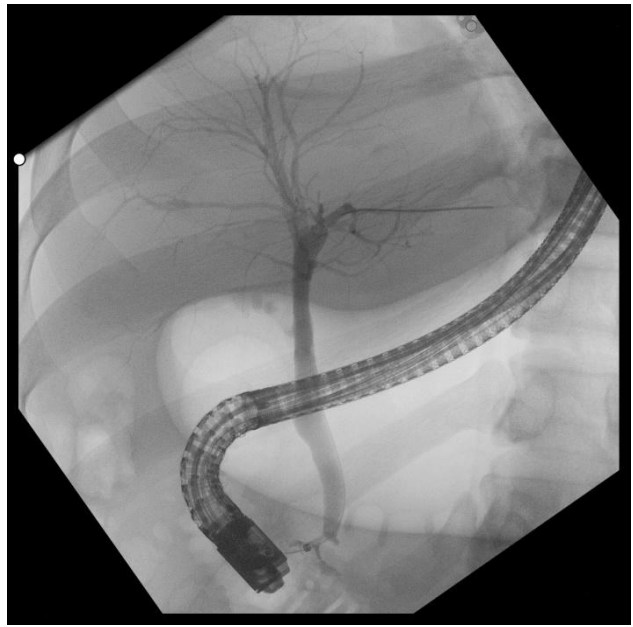

*Figure 1. ERCP image showing the endoscope, the biliary tree and the main pancreatic duct.*

## 2. Materials and Methods

*2.1. Methodology*

The methodology used for bile duct segmentation based on cholangio-MRI images, as well as their evaluation by comparing the automated segmentations with the manual segmentations (ground truth) can be seen in "Figure 2".

The Cholangio-MRI sequence images undergo preprocessing, including contrast enhancement to homogenize the images and enhance the signal-to-noise ratio. Subsequently, segmentation is performed using three techniques: thresholding via 3D Slicer, a quasi-manual method involving a defined intensity threshold; flood filling via 3D Slicer, a semi-automated approach; and region growing, another classical semi-automated method. These methods enable the delineation of regions of interest. Post-processing involves removing undesired anatomical elements from the segmented regions to enhance the outcomes. In parallel, manual segmentation is carried out directly on the Cholangio-MRI images using 3D Slicer to establish a ground truth reference. Based on this ground truth, the results of the three aforementioned segmentation methods are quantitatively assessed using measures and metrics, as well as qualitatively evaluated by experts who assess the accuracy and relevance of the segmentations.

*2.2. Manual and calculated segmentations*

In order to quantify the performance of the segmentation methods, a robust ground truth via manual segmentation is essential. Manual segmentation of all cases was performed, slice by slice, by gastroenterologists using the 3D Slicer software (Version 4.13). Manual segmentation is very time-consuming, taking between 20 and 30 hours per model. These segmentations were subsequently subjected to review and validation by an expert radiologist, who verified the absence of outliers, false communicating IHDs[*], and false non-communicating IHDs[*] [3]. For each sequence, three calculated segmentations were generated. First, two segmentations based on semi-automatic tools were performed with the 3D Slicer software: a segmentation based on the thresholding of pixels according to their intensity, and a segmentation based on flood filling, a classical algorithm that groups together a related set of pixels of the same intensity delimited by contours. Then, a third segmentation based on an automatic method of the region growing type developed internally at BMBI laboratory, was performed. This type of segmentation starts with a seed point or region and iteratively adds neighboring pixels or regions that have similar properties.

We chose to segment a majority of challenging ERCP cases with malignant hilar strictures, for which gastroenterologists may not achieve optimal results, especially in the case of hilar cholangiocarcinoma (CCK) [3]. For these patients, the manual segmentation step took approximately 200 to 300 hours of work.

[*] IHDs (intrahepatic ducts)

*2.3. Dataset*

Cholangio-MRI images of patients treated in the digestive endoscopy department of Saint-Antoine Hospital were collected in a DICOM format and de-identified. This study responds to MR004 guidelines in France on retrospective studies of anonymized data and did not require ethics committee approval.

Cholangio-MRI images are based on the bile hypersignal in T2 weighted MRI sequences, characterized by a short TE (700 to 884 ms) and a long TR (3500 to 4000 ms), and cancelling out of the surrounding tissue signal in order to retain the bile signal only. These sequences therefore allow (without using contrast medium) visualization of the bile ducts. The image sequences are acquired in the coronal plane. Ten patients were included for analysis, two with normal bile ducts and eight with malignant stenosis at the level of the hilum.

Imaging was performed on a GE Medical Systems scanner, Optima MR 450 w, with a magnetic field of 1.5 T. The accuracy of segmentation strongly depends on the spatial resolution of the images. In our case, the size of voxel was between (1.094 x 1.094 x 1.5 $mm^3$) and (0.664 x 0.664 x 2 $mm^3$).

*2.3. Data preprocessing*

Cholangio-MRI images have variable characteristics (contrast, noise, average intensity value). Pre-processing thus plays a fundamental role in the performance of image segmentation algorithms in terms of capacity and visual quality [9].

In order to facilitate the segmentation, the contrast between the structure of interest (i.e., bile duct) and the surrounding structures (background, surrounding organs) was maximized in two different ways [10]: manually on the 3D Slicer software for the thresholding and flood filling methods, and automatically for the region growing method using dynamic cropping (this technique was compared to Wiener denoising filter and showed better results).

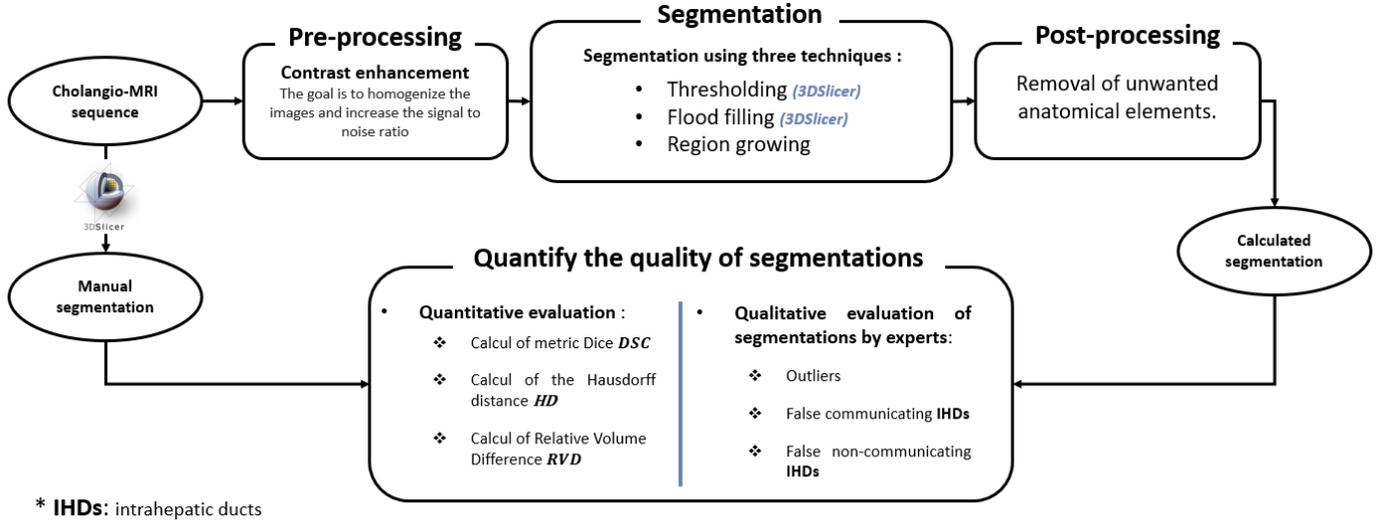

*Figure 2. Automated Segmentation of Bile Ducts from Cholangio-MRI Images and Comparative Evaluation*

## 2.4. Segmentation methods

### 2.4.1. Thresholding

The thresholding method is a well-known technique in the field of medical image segmentation [11]. This technique relies on the definition of a threshold value T such that the threshold image $g$ is obtained by:

$$g(x,y) = \begin{cases} 1 \text{ if } f(x,y) > T \\ 0 \text{ if } f(x,y) \leq T \end{cases} \quad (1)$$

Where:

$x, y$ : the coordinates of the pixel.

$f(x, y)$ : the gray level of the pixel in the image $f$.

The global thresholding technique is mainly used when the intensity distribution between the foreground and background structures is very distinct. In that case, a single threshold value can simply be used to differentiate the two structures. This is not the case in cholangio-MRI images however [11]. Hence, we decided to use two different thresholds $T_{min}$ and $T_{max}$, allowing the definition of a range of intensity to be segmented.
In this case the threshold image $g(x, y)$ is defined by:

$$g(x,y) = \begin{cases} 1 \text{ if } T_{min} < f(x,y) \leq T_{max} \\ 0 \text{ if not} \end{cases} \quad (2)$$

In order to achieve a fully segmented biliary tree while minimizing unwanted elements, the $T_{min}$ and $T_{max}$ windowing values were manually and carefully selected in 3D Slicer, varying from one sequence to another and even from one slice to another.

### 2.4.2. Flood filling

Flood filling algorithm is a region-approach segmentation technique which leads to fill a region of an image with a given grayscale value [12]. The principle of the flood filling algorithm for grayscale image segmentation follows the following steps:
- Select a starting point on the image.
- Specify a grayscale value for the desired fill.
- Set a threshold tolerance to define the grayscale range for filling. This range determines neighboring values within which filling will occur.

- Add the starting point to a stack.
- While the stack is not empty, repeat the following steps:
  - Retrieve the first element of the stack and pop it.
  - If the current point is within the image and its grayscale intensity is within the tolerance range, change its grayscale intensity to the target fill color.
  - Add to the stack the neighboring pixels that have not yet been filled and whose grayscale color is within the tolerance range.

### 2.4.3. Region growing

The region growing method is a region-approach segmentation technique which can be broken down into two steps:
- The selection in the image of one or more priming areas, often called "seeds" and serving as starting points for the next step. The selection of these areas can be automatic, using maximum or minimum light intensity for example, but it is also possible to ask the user to directly select areas of the image.
- The next step is a step of growing of the starting regions: the idea is to gradually grow the seed regions by aggregating the pixels that meet a criterion, called the homogeneity criterion. This criterion ensures that the same region grows by aggregating pixels with similar characteristics.

It should be noted that only one priming zone is required to segment all the images within a cholangio-MRI by relying on propagation.

After comparing several aggregation criteria, (Otsu threshold, detector of Canny, and fixed threshold, Sauvola's threshold), we chose the Sauvola's threshold because it lead to better segmentation [11]. The Sauvola threshold, is widely used as a local thresholding method [13]. The Sauvola approach calculates the threshold for each pixel based on the mean and standard deviation of pixel intensities within a local neighborhood. The equation for the Sauvola threshold is defined as follows:

$$T_{Sauvola}(x,y) = Mean(I_{Z3*3}(x,y)) * \left(1 + k * ((std(I_{Z3*3}(x,y))/R - 1)\right)$$
(3)

In this equation, $T_{Sauvola}(x,y)$ represents the computed threshold for the pixel at position $(x,y)$ in the image. $I_{Z3*3}(x,y)$ denotes the set of intensity coefficients of pixels within a local 3x3 pixel area centered around pixel $(x,y)$. The parameters $k$ and $R$ play key roles in this equation:

- $k$ is a sensitivity parameter that controls the influence of local variation on thresholding. A higher value of $k$ makes the thresholding more stringent, while a lower value makes it more tolerant.
- $R$ is a scale parameter that adjusts the standard deviation. Increasing it gives more consideration to the local standard deviation in the threshold calculation.

By adjusting these parameters, including using $k = 0.3$ and $R = 100$ for cholangio-IRM images, we can fine-tune the Sauvola threshold to better align with the specific characteristics of the image and local intensity variations.

### 2.5. Evaluation criteria

### 2.5.1. Quantitative evaluation criteria

In this paper we used three evaluation indicators; the Dice similarity coefficient (DSC), the Hausdorff distance (HD) and the Relative Volume Difference (RVD) to measure the accuracy and quality of the three segmentations generated.

Given $X$ a set of pixels belonging to a ground truth, and $Y$, a set of pixels belonging to a segmented object, the Dice index is defined as follows [14]:

$$DSC(X,Y) = \frac{2|X \cap Y|}{|X|+|Y|}$$
(4)

$HD$ is an indicator of the largest segmentation error. As illustrated in "Figure 3", for two point sets $X$ and $Y$, the one-sided $HD$ from $X$ to $Y$ is defined as [15]:

$$HD(X,Y) = max_{x \in X} \, min_{y \in Y} \, \|x - y\|_2$$
(5)

and similarly:

$$HD(Y,X) = max_{y \in Y}\ min_{x \in X}\ \|x - y\|_2 \quad (6)$$

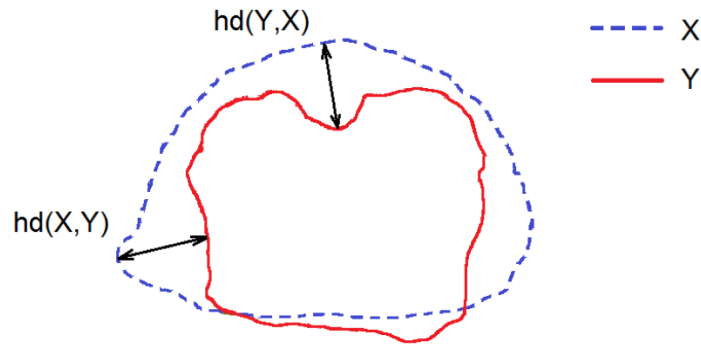

*Figure 3. A schematic showing the hausdorff distance between points sets X and Y [15].*

Finally, the Relative Volume Difference ($RDV$) is calculated by taking the absolute difference between the volume of the automatically generated segmentation X and the manually generated segmentation Y, divided by the volume of the manual segmentation Y [16]. $RDV$ is defined as:

$$RVD = \frac{|X-Y|}{Y} \quad (7)$$

*2.5.2. Qualitative evaluation criteria*

In order to assess the quality of the segmentations generated by the three different methods, a qualitative evaluation was also performed by three expert gastroenterologists. The evaluators first assessed the similarity in caliber and completeness of the common bile duct and intrahepatic bile ducts. Then, they visually counted the number of outliers, false communicating IHDs, and false non-communicating IHDs in the segmentations, as shown in "Figure 4, 5 and 6" respectively. The evaluators were blinded to the segmentation method used and asked to reach a consensus agreement for each case and each segmentation. The manual segmentation was defined as the gold standard for these evaluations.

In the context of bile duct segmentation, outliers typically refer to segmentation errors in which parts of the bile duct structure are either incorrectly segmented or are missed altogether "Figure 4". This can occur due to a variety of factors, such as variations in image quality, imaging artefacts, or limitations in the segmentation algorithm.

False communicating IHDs in bile duct segmentation refer to segmentation errors in which regions of the image that are not part of the bile duct structure are included in the segmentation and connected to the bile ducts "Figure 5". These errors can lead to the incorrect connection of segments in the biliary tree that are not actually connected, resulting in a potentially misleading representation of the biliary anatomy. False connections can occur due to various factors, such as differences in image quality, imaging artefacts, or limitations in the segmentation algorithm. In particular, anatomical structures in the vicinity of the bile ducts that have similar signal intensities can contribute to the occurrence of false connections.

False non-communicating IHDs occur when two adjacent segments of the bile duct are mistakenly identified as distinct entities, resulting in the erroneous division of a single segment into two separate parts "Figure 6".

These types of segmentation error can cause confusion and misinterpretation of the anatomy of the bile duct, potentially leading to incorrect medical decisions. As such, avoiding outliers, false communicating IHDs and false non-communicating IHDs is an important goal in the accurate segmentation of the bile duct.

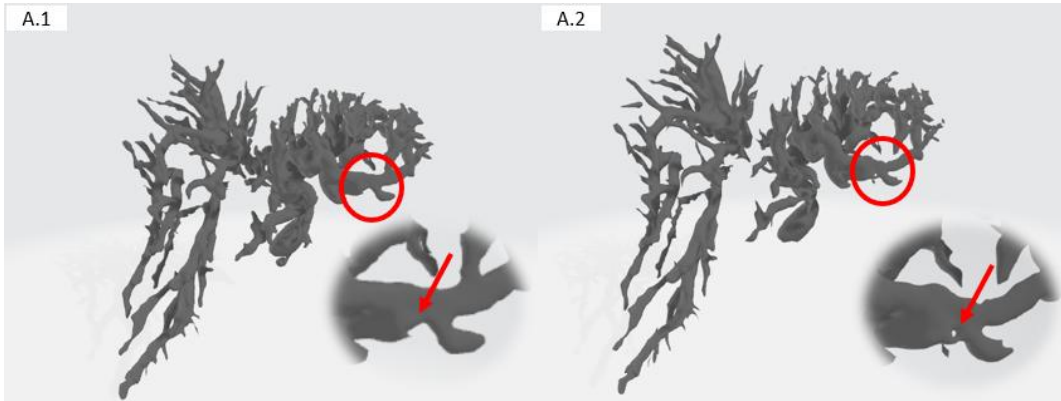

*Figure 4.* Example of outlier, Ground truth (A.1), Calculated segmentation (A.2).

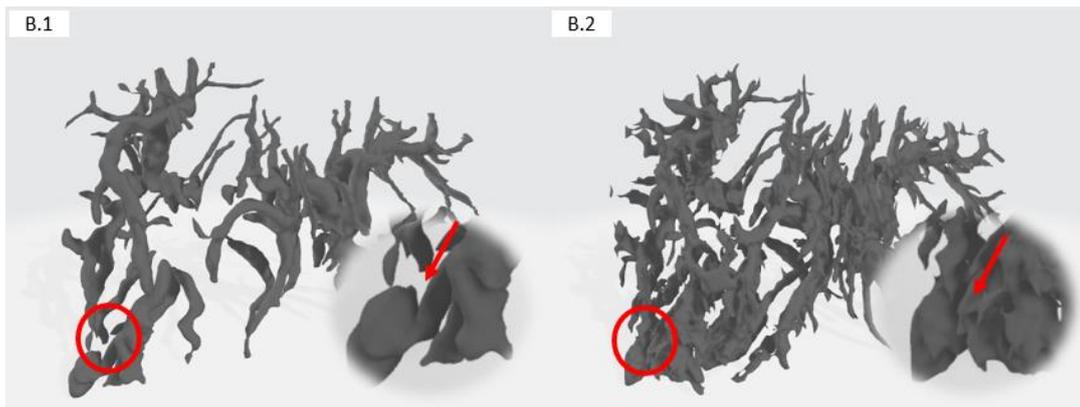

*Figure 5.* Example of false communicating IHDs, Ground truth (B.1), Calculated segmentation (B.2).

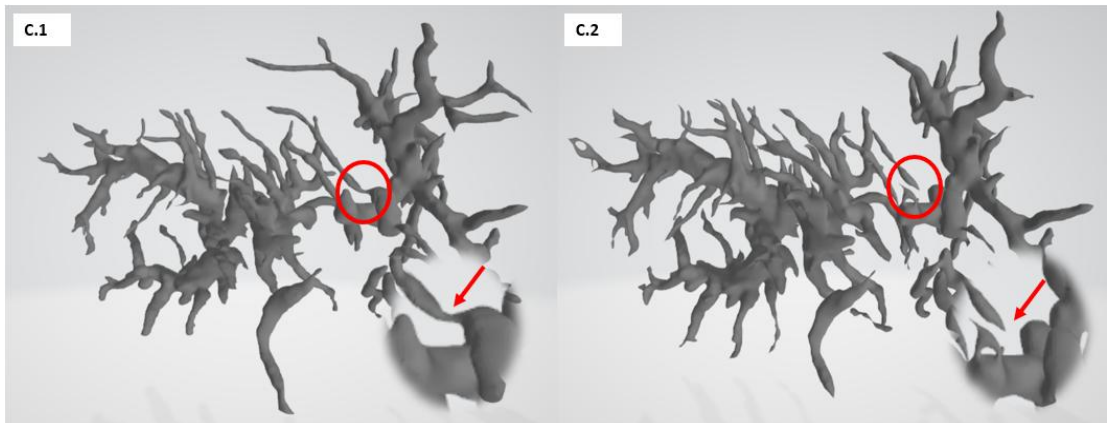

*Figure 6.* Example of False non-communicating IHDs, Ground truth (C.1), Calculated segmentation (C.2).

## *2.6. Statistical analysis*

A statistical analysis was conducted to determine if there were any significant differences between the techniques. This analysis was performed using the GraphPad Prism 8.4.3 software (GraphPad, San Diego, CA, USA). A one-way ANOVA test was employed. The results were considered statistically significant if the p-value was less than 0.05, which is a common threshold for determining significance. In the figures, any data with a p-value less than 0.05 were highlighted with an asterisk to indicate their significance.

# 3. Experimental Results

Automated segmentation using the three methods (thresholding, flood filling, and region growing) was feasible in all cases "Figure 7". The mean DSC values were 0.819 ± 0.057, 0.772 ± 0.059, and 0.657 ± 0.110 for thresholding, flood filling, and region growing, respectively. The HD were 0.816mm ± 0.353, 1.291mm ± 0.656, and 1.963mm ± 1.091 for thresholding, flood filling, and region growing, respectively. The RDV values were 0.188 ± 0.118, 0.235 ± 0.172, and 0.539 ± 0.349 for thresholding, flood filling, and region growing, respectively.

The number of outliers for the three methods were 6.2 ± 3.86, 8.2 ± 3.15, and 7.9 ± 3.75. The number of communicating IHDs switched to non-communicating IDHs were 2.0 ± 2.00, 2.1 ± 1.57, and 4.8 ± 6.24 for thresholding, flood filling, and region growing, respectively. Finally, the number of non-communicating IDHs switched to communicating IDHs were 0.2 ± 0.40, 0.0 ± 0.00, and 0.1 ± 0.30 for thresholding, flood filling, and region growing, respectively. These results indicate that thresholding had the highest mean DSC value, the lowest HD, and the lowest RDV, while all three methods had a large number of communicating IHDs switched to non-communicating and a large number of outliers. These results can be seen in table 1.

The Thresholding method exhibited a significantly better performances (quantitative and qualitative criteria) (p=0.05) in bile duct segmentation compared to the Region growing method "Figure 8". This is due to the possibility of manually choosing the threshold for each area in the biliary tree (CBD, HH, and IHD) (which is not reproducible) as opposed to the other two methods. Except for the DSC, there is no significant difference between the two region-based approach methods (Flood filling and Region growing). However, none of the three methods are robust for bile duct segmentation, as evidenced by a Dice score < 0.82 and a Hausdorff distance > 0.816.

| Segmentation method | Quantitative evaluation criteria | | | Qualitative evaluation criteria | | |
|---|---|---|---|---|---|---|
| | *DSC* | *HD (mm)* | *RDV* | *Outliers* | *false communicating IHDs* | *false non-communicating IHDs* |
| Thresholding | **0.819 ±0.057** | **0.816 ±0.353** | **0.188 ±0.118** | **6.2 ±3.86** | **2.0 ±2.00** | 0.2 ±0.40 |
| Flood filling | 0.772 ±0.059 | 1.291 ±0.656 | 0.235 ±0.172 | 8.2 ±3.15 | 2.1 ±1.57 | **0.0 ±0.00** |
| Region growing | 0.657 ±0.110 | 1.963 ±1.091 | 0.539 ±0.349 | 7.9 ±3.75 | 4.8 ±6.24 | 0.1 ±0.30 |

*Table 1. Experimental results of bile duct segmentation*

**DSC:** *Dice*, **HD:** *Hausdorff distance*, **RDV:** *Relative Volume Difference*, **IHDs:** *Intrahepatic ducts*

[Figure 7: Table comparing Ground truth, Thresholding, Flood filling, and Region growing segmentation methods across Slice 1, Slice 2, Slice 3, and 3D Reconstruction]

*Figure 7.* Comparison of bile duct segmentation for the three methods with ground truth (bleu box is labeled CBD, yellow box is labeled HH, red box is labeled IHD, white circle: example of over-segmentation, orange circle: example of under-segmentation), CBD: Common bile duct, HH: Hepatic hilum, IHDs: Intrahepatic ducts.

## 4. Discussion and Future work

We will initiate the discussion by contextualizing our findings within the existing body of knowledge. Understanding the intricate anatomy of the bile ducts is crucial for the success of endoscopic retrograde cholangiopancreatography (ERCP). However, its delicate nature, coupled with interindividual anatomical variations, poses significant challenges. Although preoperative imaging techniques such as abdominal CT and MRCP, along with intraoperative 2D fluoroscopy, provide valuable insights, they lack a comprehensive overview. In this context, the segmentation and 3D reconstruction of the bile ducts from cholangio-MRI images offer a promising solution to address this limitation [3]. Yet, manual segmentation, despite its robustness, is time-consuming, underscoring the need for automated solutions to enhance diagnostic accuracy and procedural efficiency [4], while also reducing X-ray exposure for both patients and physicians.

Despite the potential advantages, automated cholangio-MRI segmentation faces various challenges, including image brightness disparities, low spatial resolution, variable image quality, and the inability to detect bile duct walls in cases of stenosis. Previous research has explored techniques such as multiscale vesselness filters [7], phase symmetry filters [6], and active contour methods [8], but none have provided a fully robust solution.

Our study rigorously evaluated classical segmentation methods, unveiling their shortcomings in cholangio-MRI image segmentation. DSC values, HD measurements, and RDV values illuminated the limitations of thresholding, flood filling, and region growing methods. Notably, none of these methods achieved the desired robustness, indicating a critical need for innovative alternatives.

However, upon a more comprehensive analysis of the results, we were able to identify specific areas where these methods exhibited tendencies toward over-segmentation or under-segmentation of the regions of interest. For instance, in certain regions with indistinct boundaries, the region growing method was prone to incorporating surrounding pixels that were not inherently part of the region in question, thereby leading to a slight over-segmentation (identified by a white circle in "Figure 7"). Conversely, in regions characterized by low contrasts, the Flood filling method occasionally omitted portions of the region, resulting in under-segmentation (identified by an orange circle in "Figure 7"). These limitations (under/over segmentations) result in false communicating IHDs and false non-communicating IHDs in the 3D reconstruction.

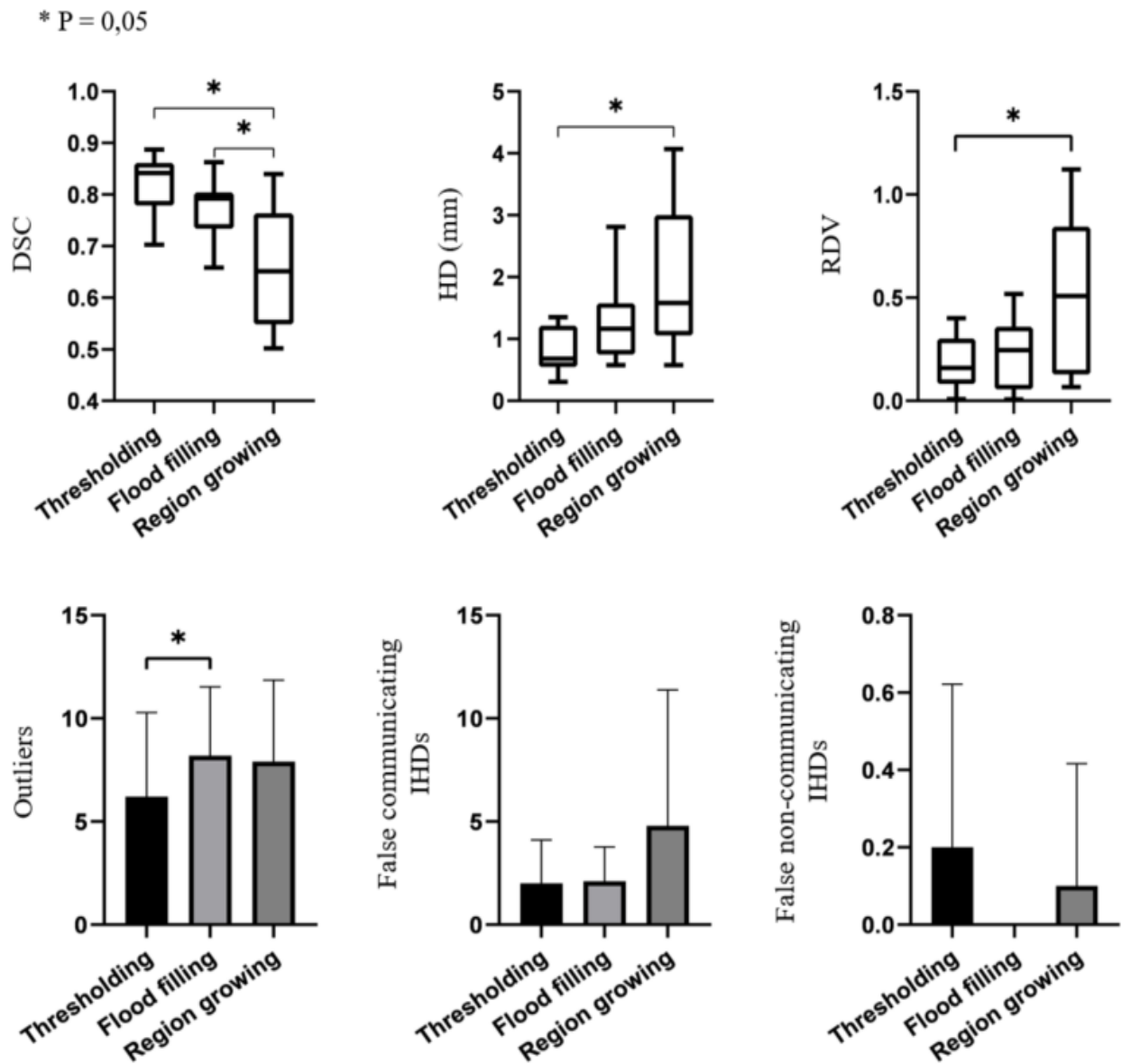

*Figure 8. Graphs of DSC, HD, RDV, Outliers, False communicating IHDs and False non-communicating IHDs for the three segmentation methods.*

A qualitative evaluation of the segmentations by three expert endoscopists was conducted. The results are presented in "Table 1". This qualitative evaluation of classical bile duct segmentation methods revealed several drawbacks. Firstly, these methods tend to create false connections between intrahepatic bile ducts, potentially leading to misinterpretation and incorrect treatment decisions during CPRE, with exposure to infectious risks, with a particularly high risk of prosthetic misplacement. Secondly, these methods fail to provide a complete segmentation of the biliary tree, resulting in gaps that hinder practical use by endoscopists. Other less significant limitations of these classical segmentation methods include over-segmentation and the presence of outliers.

To address these gaps, we are currently leveraging and developing a U-Net-based model for precise cholangio-MRI segmentation. Our approach yielded promising results, confirming its effectiveness in terms of segmentation accuracy and quality. While these results are encouraging, further work is essential to consolidate and generalize our conclusions across a broader patient cohort. We intend to extend the application of our model to various hepatobiliary conditions, evaluating its adaptability and robustness across diverse clinical scenarios.

Future improvements to our model encompass exploring U-Net architecture variations and integrating advanced deep learning techniques such as pre-trained networks and attention mechanisms. These endeavors aim to enhance segmentation performance and result quality. Concurrently, our data augmentation pipeline will be refined to improve model generalization, accommodating the inherent clinical variability in cholangio-MRI images.

## 5. Conclusion

In this study, three classical techniques used for bile duct segmentation based on cholangio-MRI images were evaluated. Our results suggest that these methods are not suitable for clinical practice, as they fail to provide perfectly accurate results. Given that manual segmentation is time consuming, development of a rapid, automated, reproducible and accurate alternative method is warranted. According to our preliminary findings, deep learning techniques exploiting a large number of cholangio-MRI images may provide a viable solution. This approach may overcome the limitations of conventional methods. The development of a new method based on deep learning techniques is currently being conducted by our team.

## Acknowledgements

This work is a component of a continuous research initiative that involves collaboration between the Saint-Antoine and Mondor Hospital of the AP-HP, the Institute of Intelligent Systems and Robotics (ISIR), and the Laboratory of Biomechanics and Bioengineering (BMBI). The principal aim of this initiative is to enhance and ensure the overall safety of ERCP. We would like to extend our sincere gratitude to the Labex CAMI (ANR-11-LABX-0004), which provided essential support for this work through the Investissements d'Avenir program of the French Government.